# Analyzing the State of COVID-19: Real-time Visual Data Analysis, Short-Term Forecasting, and Risk Factor Identification

Jiawei Long [1*]

[1] *Department of Biostatistics, UCLA Fielding School of Public Health*
*University of California, Los Angeles, United States*

***Abstract*** - The COVID-19 outbreak was initially reported in Wuhan, China, and it has been declared as a Public Health Emergency of International Concern (PHEIC) on 30 January 2020 by WHO. It has now spread to over 180 countries, and it has gradually evolved into a world-wide pandemic, endangering the state of global public health and becoming a serious threat to the global community. To combat and prevent the spread of the disease, all individuals should be well-informed of the rapidly changing state of COVID-19. To accomplish this objective, I have built a website to analyze and deliver the latest state of the disease and relevant analytical insights. The website is designed to cater to the general audience, and it aims to communicate insights through various straightforward and concise data visualizations that are supported by sound statistical methods, accurate data modeling, state-of-the-art natural language processing techniques, and reliable data sources. This paper discusses the major methodologies which are utilized to generate the insights displayed on the website, which include an automatic data ingestion pipeline, normalization techniques, moving average computation, ARIMA time-series forecasting, and logistic regression models. In addition, the paper highlights key discoveries that have been derived in regard to COVID-19 using the methodologies.

## 1- Introduction

COVID-19 is an infectious disease caused by severe acute respiratory syndrome coronavirus. It was first identified in December 2019 in Wuhan, China, and has resulted in an ongoing pandemic. The virus is typically rapidly spread from one person to another via respiratory droplets produced during coughing and sneezing. It is considered most contagious when people are symptomatic, although transmission may be possible before symptoms show in patients. Time from exposure and symptom onset is generally between two and 14 days, with an average of five days.[1] Since knowledge about this virus is rapidly evolving due to its rapid spread and uncertain mutations, the public are urged learn about about the most recent state of the virus on a regular basis in order to stay informed. To contribute to the fight against COVID-19, I have created a COVID-19 real-time tracker website[2] to serve as a platform to provide the latest status of the spread of the disease and to present useful analytical insights of the disease. It includes features such as odometers of the latest state of COVID-19 cases, trend analysis, and short-term forecast in 181 different countries, informative visualizations of the most common symptoms and risk factors, and patient demographic distributions. The website retrieves the most recent data from reliable data sources on an hourly basis and transforms the data into informative visualizations and analytical insights.

Section 2 of the paper discusses the related work in the realm of data analysis of COVID-19 and highlights the innovative differences between the study presneted in this paper and the previous studies.

Section 3 to 6 of the paper provide a brief description of every feature of the website and discuss relevant methodologies behind each feature. Each section begins with an overview subsection to introduce the components and functions of the feature, followed by more detailed discussions of the methodologies that are utilized to build the feature and key insights from the relevant analytic study. In section 3, the paper discusses the mechanism of real-time data retrieval and various data processing techniques that are utilized to generate the insights in the *Overview* feature of the website. Section 4 discusses the concepts of moving average and ARIMA time-series forecasting, and how they are implemented in the *Trend* feature of the website. In section 5, the paper discusses n-gram tokenization, min-max normalization, and logistic regression model for studying common symptoms from the COVID-19 patients and risk factors that potentially increase the patient's likelihood of dying from the disease. Section 6 highlights the demographic distributions of infected patients and explains the applications of hypothesis testing in discovering potential differences in demographic distributions of different patient groups. Lastly, section 7 concludes the paper with major components of this study and discusses the future of the research and development regarding COVD-19.

## 2- Related Work

Since the initial outbreak of COVID-19, there has been a number of attempts to analyze the state of the virus from the perspective of data analytics. In early 2020, Samrat K. Dey released one of the earliest papers that analyzes the outbreak of COVID-19 through visual exploratory analysis, focusing on the spatial component and comparing the spread in the Hubei province, other Chinese provinces, and the rest of the world.[2] The author has also published an interactive notebook that consists of various interactive visualizations on a website. While the website serves as a great platform for the audience to gain deeper insights, the website's notebook remains in a static state and does not update as the state of COVID-19 quickly changes with time. This inspired me to present my analysis of COVID-19 in a highly reproducible manner, in which case all data ingestions and analysis procedures can be repeated automatically and persistently. In addition to visual exploratory analysis, there was also significant effort in understanding and forecasting the spread of the virus. In an early paper presented by Yit C. Tong, the author attempts to model and project the spread of the virus through mathematical analysis that utilizes moving average and the SIR model, a traditional epidemic model.[3] Similar work was also conducted in a study led by Baoquan Chen, in which case the author attempts to forecast the future trend of COVID-19 through the C-SEIR model, an extension of the traditional SIR model.[4] After reproducing some of the previously mentioned work, I have found epidemic modeling to be limiting in terms of modeling the uncertainty of COVID-19 due to its strict model assumptions that conflict with the spread of COVID-19. Thus, I have taken a different direction to model the state of COVID-19 with a more flexible framework that utilizes time-series modeling with moving average in my study. Moreover, There were many clinical studies on discovering the common symptoms and identifying potential risk factors associated with the virus through traditional medical research methods such as experimental design and analysis.[5][6][7] In this paper, I have proposed an innovative way to analyze medical records retrospectively to estimate the relative prevalence of various symptoms and identify potential risk factors through natural language processing techniques. This paper also significantly differs from the existing related work in the field because it presents a larger breadth of analysis on various aspects of COVID-19, ranging from visualizations of its latest state to statistical modeling that produces short-terming forecasting and identifies potential risk factors.

# 3- Feature 1: Overview

## *3-1- Introduction*

The landing page of the website is the *Overview* page, and it presents the most current states of COVID-19 at a global scale (Figure 1). The top of the page has three odometer boxes to display the total confirmed cases, the total deaths, and the total recovered cases along with their respective daily new counts. The bottom half of the page contains a user-interactive control panel and a display window. The users are able to apply population normalization or log transformation to the visualizations in the display window through the widgets in the control panel.

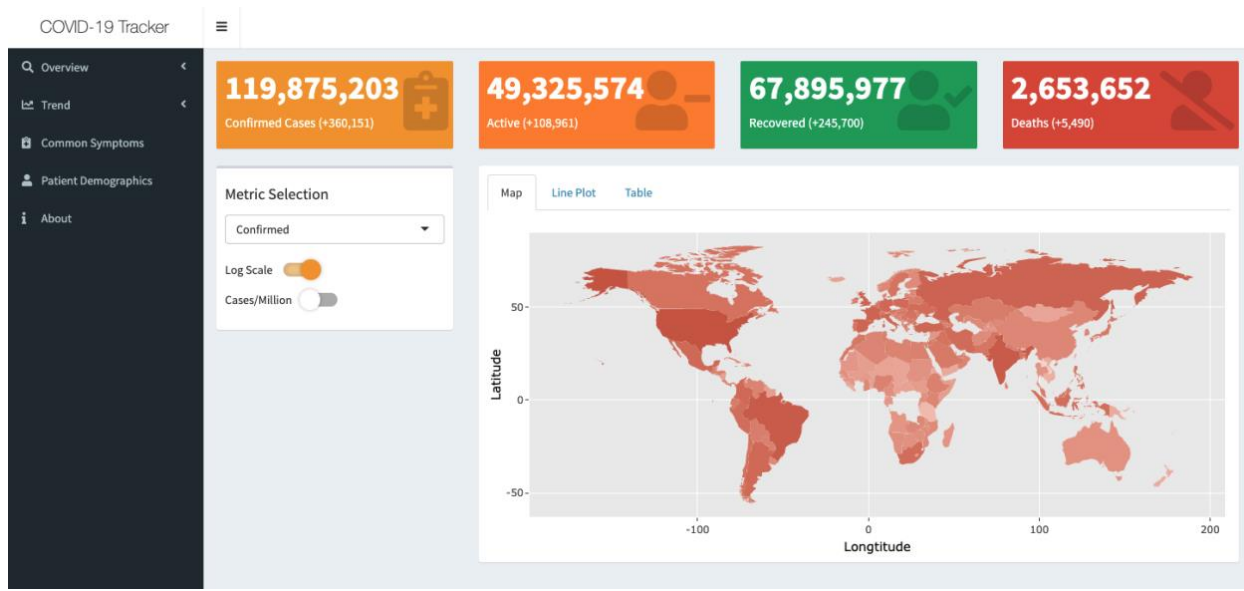

Figure 1. Overview Page (Website Landing Page).

The display window has the viewing options of heat map visualizations (Figure 2), time-series line plots (Figure 3), or data tables (Figure 4). All visualizations have interactive features such as tooltip and zooming, and all tables can be interactively sorted by clicking on column names. The heat map visualization allows the users to quickly assess the severity of COVID-19 in different geographical locations while the time-series line plot shows a comparison of the most affected regions on a standardized time scale. The data table provides the users with the flexibility to explore and search for data of their interest.

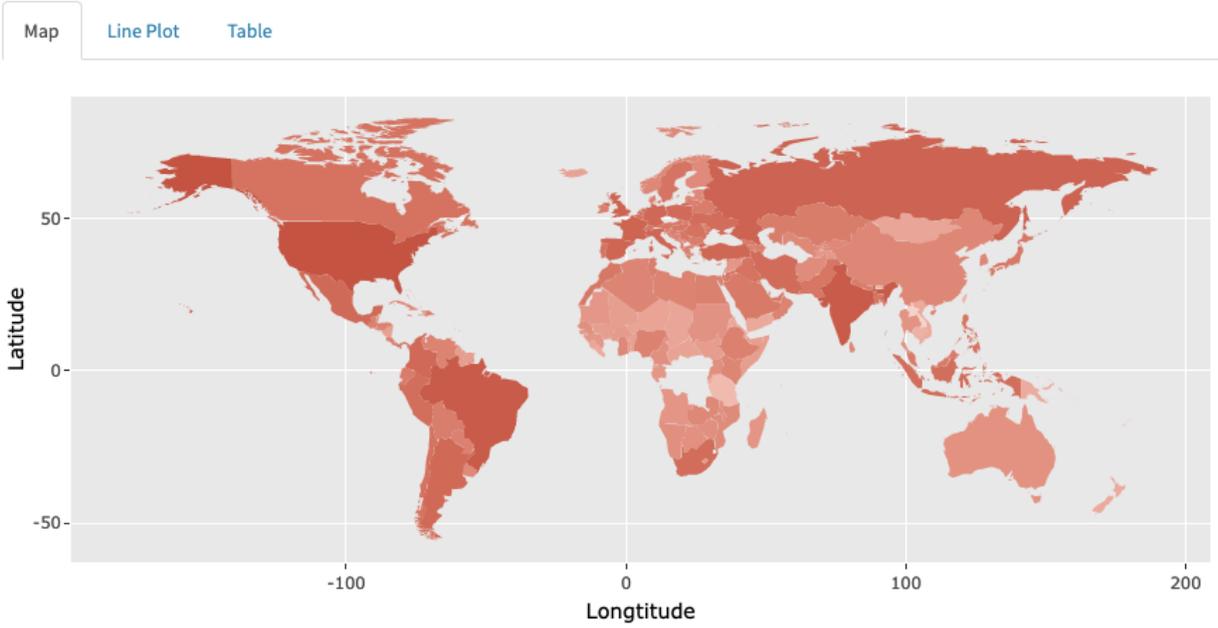

Figure 2. Global Confirmed Cases Heat Map.

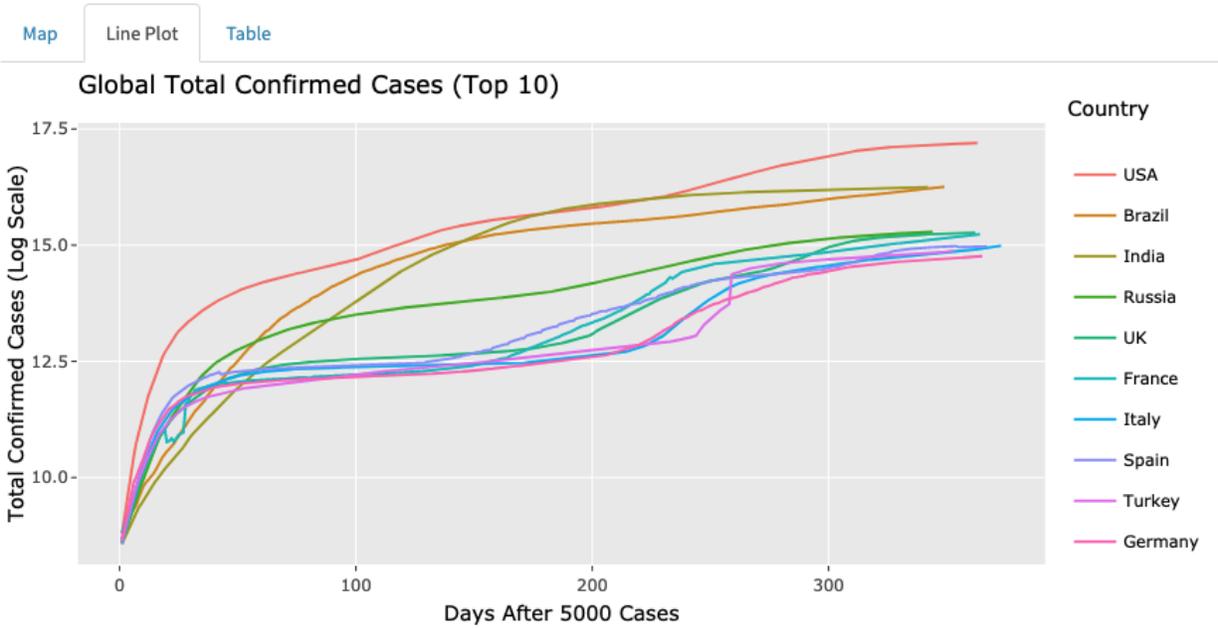

Figure 3. Global Confirmed Cases Time-Series Line Plot.

| | Country | Confirmed | Deaths | Recovered | Population | Confirmed/M | Deaths/M | Recovered/M |
|---|---------|-----------|--------|-----------|------------|-------------|----------|-------------|
| 1 | USA | 29438775 | 534888 | | 330610570 | 89043.66 | 1617.88 | 1617.88 |
| 2 | Brazil | 11483370 | 278229 | 10113487 | 212253150 | 54102.24 | 1310.84 | 1310.84 |
| 3 | India | 11385339 | 158725 | 11007352 | 1377233523 | 8266.82 | 115.25 | 115.25 |
| 4 | Russia | 4341381 | 90558 | 3949891 | 145922010 | 29751.38 | 620.59 | 620.59 |
| 5 | UK | 4271710 | 125753 | 11972 | 67814098 | 62991.47 | 1854.38 | 1854.38 |
| 6 | France | 4131874 | 90583 | 279752 | 65244628 | 63328.95 | 1388.36 | 1388.36 |
| 7 | Italy | 3223142 | 102145 | 2589731 | 60479424 | 53293.2 | 1688.92 | 1688.92 |
| 8 | Spain | 3183704 | 72258 | 150376 | 46751175 | 68098.91 | 1545.59 | 1545.59 |
| 9 | Turkey | 2879390 | 29489 | 2701076 | 84153250 | 34216.03 | 350.42 | 350.42 |
| 10 | Germany | 2578842 | 73463 | 2368995 | 83730223 | 30799.42 | 877.38 | 877.38 |

Figure 4. Global Data Table for COVID-19 Statistics by Country.

The purpose of this feature is to provide the audience with a concise view of the severity of COVID-19 in different locations and inform the audience of the latest progression of the disease easily. The options of applying log transformation and population normalization allow the audience to observe the state of COVID-19 from different perspectives while the interactive table allows the audience to explore specific metrics of their interest. These options are useful because they allow users to access more detailed information under a different context. The website also consists of an overview page specific for the United States at a state level, similar to the shown page corresponding to a global scale at the country level. The overview page for United Stats shares the same features discussed above and it can be accessed by expanding the *Overview* feature and clicking on the *U.S.* tab (Figure 5). The subsequent subsections discuss the mechanism of automatic data ingestion pipeline and various data processing techniques that are utilized to generate the insights in the Overview feature of the website.

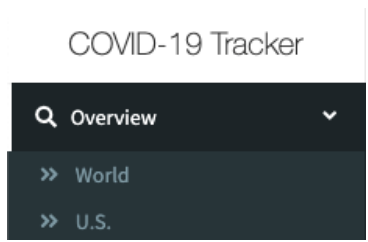

Figure 5. Two Subpages of the Overview Feature.

*3-2- Automatic Data Ingestion Pipeline*

      To ensure the accuracy and the reliability of the website's content, the website's server retrieves the newest data from the COVID-19 data repository maintained by the Center for Systems Science and Engineering at Johns Hopkins University on an hourly basis. The data repository is regulated by the Johns Hopkins University Center for Systems Science and Engineering and supported by the ESRI Living Atlas Team and the Johns Hopkins University Applied Physics Lab. The data source is supposed to be updated numerous times throughout the day, and the validity of the data is verified by researchers at Johns Hopkins University. The content displayed in the *Overview* feature is therefore derived from a real-time and reliable data source.

      To achieve an automatic data ingestion process, I have created a data pipeline using Apache Airflow to retrieve the most recent data from the CSSE data repository by Johns Hopkins. The data pipeline contains a protocol to download and ingest the most recent data from the data source, and it is governed by a scheduler to run at the beginning of every hour. During each data ingestion process, the pipeline's program will download the data by obtaining the current date and accessing the data source with a modified URL. For example, the newest daily report data file can always be accessed using the link "https://raw.githubusercontent.com/CSSEGISandData/COVID-19/master/csse_covid_19_data/csse_covid_19_daily_reports/MM-DD-YYYY.csv, where "MM-DD-YYYY" is used as a placeholder to store the current date. The program recognizes the date in Pacific Standard Time and inputs it into the link's placeholder when the server tries to retrieve the newest data. The link directs to a raw CSV file in HTML format, which can be directly downloaded using a pre-specified R script. Once the data file is downloaded successfully, it will be ingested and stored in an AWS S3 bucket. When the website is opened by a user, the server behind the website will read the data stored in the AWS S3 bucket and preprocess it into different data frames that support the content on the website. If the process were to be unsuccessful due to unforeseen circumstances, the web server will load up the most recent data file that it has ingested previously to support the content on the web page. Figure 6 summarizes the data ingestion process of the website.

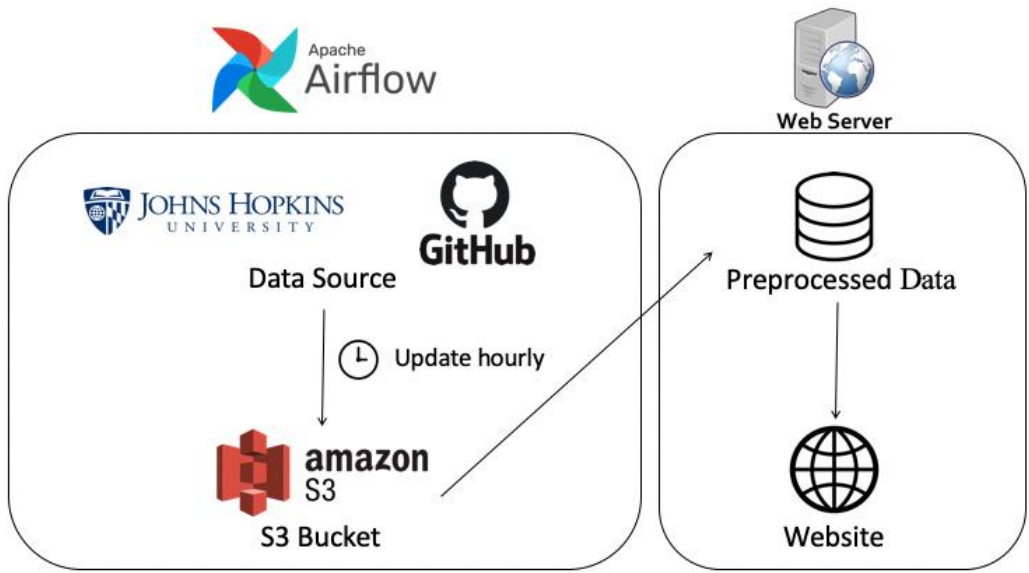

Figure 6. Automatic Data Ingestion Pipeline Summary.

## 3-3- Transformation and Normalization Techniques

The control panel on the page allows the users to apply log transformation and population normalization (i.e. cases per million) to the data, which interacts with the corresponding visualizations of the heat map and time-series line plot. When the user turns the log scale switch on, the logarithmic function with a base of 10 is used as a deterministic mathematical function to be applied to each point in a data set. That is, for every data point $x_i$, its value will be replaced by $y_i = \log_{10}(x_i)$. Such a transformation can significantly improve the interpretability and the appearance of the graph. The choice of using the logarithmic function is based on the nature of exponential growth associated with pandemic and the relatively large differences in the raw counts of cases across different locations in the later stages of a pandemic. The effect of log transformation is demonstrated in Figure 7.

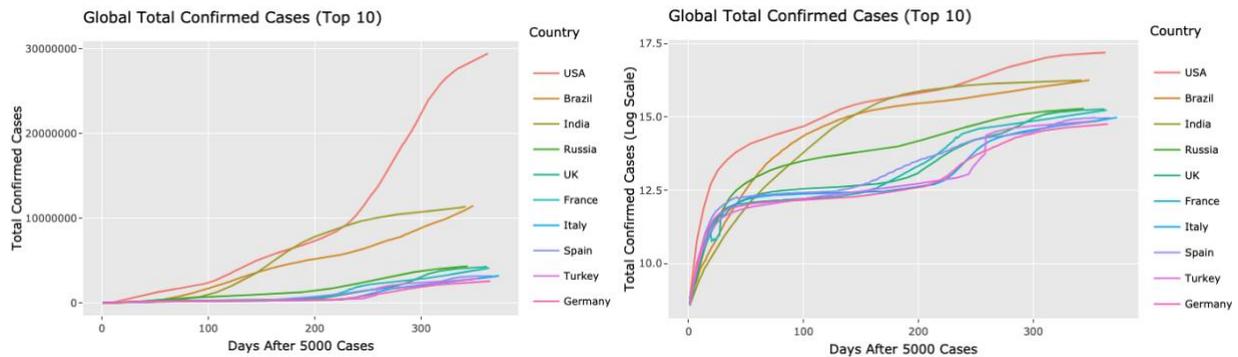

Figure 7. Before (Left) and After (Right) Log Transformation.

On the other hand, while population normalization does not necessarily improve the appearance of the visualizations, it alters the interpretation of the visualization by accounting for the population of each region. Such a perspective is beneficial because each country or state could vary significantly in its population. Assessing the number of cases per million provides a more robust estimation of the severity of the COVID-19 in each region rather than solely observing the total counts. To achieve population normalization, global country-level population data and state-level population data of the U.S. are preprocessed and stored on the server, and they will be joined to the retrieved data to produce the corresponding visualizations. Precisely, the normalization is applied in the following manner, for every data point $x_i$, its value will be replaced by $y_i = (x_i / n_j) \cdot 1{,}000{,}000$, where $n_j$ denotes the population of country$_j$. Figure 8 further displays the usefulness of the log transformation as a visual tool for studying the data.

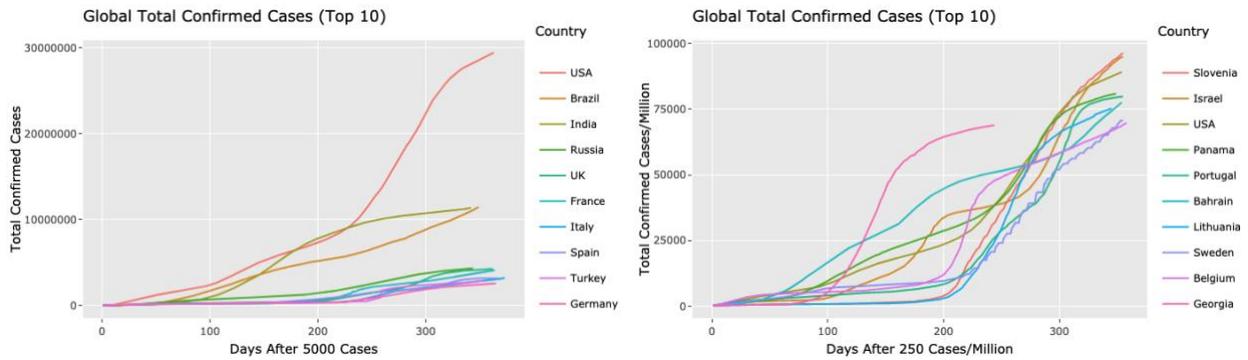

Figure 8. Before (Left) and After (Right) Population Normalization.

In addition, the time-series line plots have built-in timescale standardization. Rather than comparing the time-series data with respect to date, the plot compares them with respect to the number of days after the spread of the disease reaches a certain magnitude. Since the time frames of outbreaks are different in every region, it will be hard to compare the severity of the disease in each region in separate time frames. Hence, the application of timescale standardization helps to standardize the time-series data into a universal time scale. In conjunction with population normalization, the audience is able to compare regions that have the fastest spread of COVID-19.

## 4- Feature 2: Trend Analysis

### *4-1- Introduction*

This section discusses another useful feature of the website and its ability to display time-series data in different ways. The *Trend* feature of the website focuses on individual country-level statistics. The page contains a user-interactive control panel and a display window, where the display window shows visualizations of the daily increment of daily new cases as shown in Figures 9 and 10. In the figure, the orange bar plot represents the number of daily new cases while the black line represents the 14-days moving average of the daily new cases. The grey dotted line extended from the black line and the orange ribbon around it together represent a 14-day forecast of the number of daily new cases. The user-interactive control panel on the left allows the user to select the metric of interest (i.e. confirmed cases, deaths, or recovered cases), the country of interest (including 181 countries), and whether the scope of the plot should focus on visualizing the short-term trend or the long-term trend.

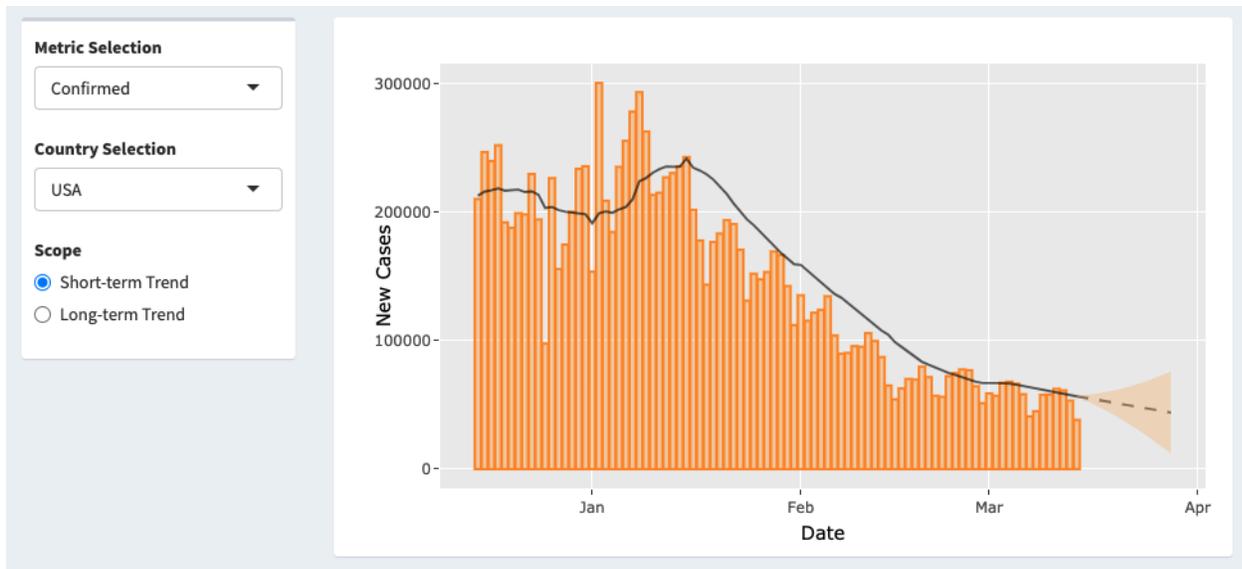

Figure 9. New Confirmed Cases Short-term Trend in USA.

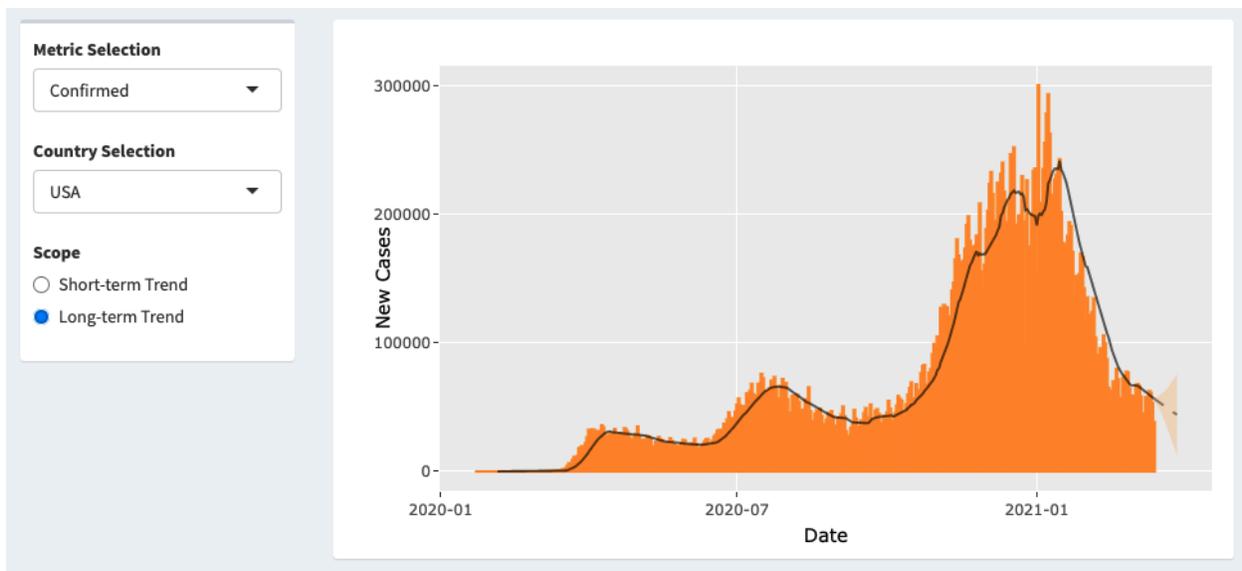

Figure 10. New Confirmed Cases Long-term Trend in USA.

The purpose of this feature is to provide insights into the trend of the spread of the disease in each country. In particular, the plot is designed to answer the pressing question of whether the curve has flattened. Since the number of daily new cases has a substantial amount of fluctuations, applying a moving average aggregation can help to reveal the underlying direction of the curve of the number of daily new cases. In addition, the moving average values display trends and patterns that can serve as the basis for time-series modeling to be used for forecasting purposes. The subsequent subsections discuss the concepts of moving average and ARIMA time-series forecasting, and how they are implemented in the *Trend* feature of the website.

## 4-2- Moving Average

One major component of the trend visualization is the moving average curve overlaid on the bar plot, and the value of the period of days used for commutating the moving average aggregation is 14 in this case. Moving average is an aggregating calculation to analyze data points by calculating a series of averages on different subsets of the full data set.[8] In this case, the method of simple moving average is used to compute the values of the moving averages. As an example, suppose we wish to calculate a simple moving average, if $X_t$ is the number of new cases at time *t*, a simple moving average at $t = m$ is computed as

$$SMA_m = \frac{X_m + X_{m-1} + X_{m-2} + \ldots + X_{m-(n-1)}}{n} = \frac{1}{n}\sum_{i=m-(n-1)}^{m} X_i \tag{1}$$

By computing a series of simple moving averages, we smooth out short-term fluctuations in the number of daily new cases and highlight longer-term trends or cycles. This is especially useful in determining the constantly changing state of the COVID-19 outbreak in a particular region.

## 4-2- ARIMA Time-Series Model

The following discussion briefly introduces the ARIMA model and provides the necessary background information for the prediction mechanism used by the website. ARIMA, short for Auto Regressive Integrated Moving Average models, are commonly used to fit time-series data using lags and the lagged forecast errors so that the fitted model can be used to forecast future values.

ARIMA models involve the notions of stationarity, differencing, autoregressive models, and moving average models.[10] An ARIMA model assumes that the input time-series data is univariate and stationary. Stationarity implies that the time series' properties are independent of the time when they were captured. Additionally, the data has a constant mean and variance. Otherwise, they need to be transformed before one can use the ARIMA model. Such a transformation process is called differencing, and the appropriate order of differencing can be determined by using methods such as the Kwiatkowski-Phillips-Schmidt-Shin (KPSS) test.[11] Differencing is a process of computing the differences between consecutive observations. It has the function of stabilizing the mean of a time series by removing changes in the level of a time series, and therefore reducing trend and seasonality. In an autoregression model, we forecast the variable of interest using a linear combination of past values of the variable. The term autoregression indicates that it is a regression of the variable against itself. Thus, an autoregressive model of order p can be written as

$$y_t = c + \phi_1 y_{t-1} + \ldots + \phi_p y_{t-p} + \varepsilon_t, \tag{2}$$

where $\varepsilon_t$ is white noise. This is like a multiple regression but with lagged values of $y_t$ as predictors. We refer to this as an *AR(p)* model, an autoregressive model of order *p*. In contrast, rather than using past values of the forecast variable in a regression, a moving average model uses past forecast errors in a regression-like model.

$$y_t = c + \theta_1 \varepsilon_{t-1} + \ldots + \theta_q \varepsilon_{t-q}, \tag{3}$$

where $\varepsilon_t$ is white noise. We refer to this as an MA(q) model, a moving average model of order *q*.

If we combine differencing with autoregression and a moving average model, we obtain an ARIM model. The full model of can be written as

$$y'_t = c + \phi_1 y'_{t-1} + \ldots + \phi_p y'_{t-p} + \theta_1 \varepsilon_{t-1} + \ldots + \theta_q \varepsilon_{t-q} + \varepsilon_t, \qquad (4)$$

where $y_t'$ is the differenced series, $p$ is the order of the autoregressive part, $d$ is the degree of differencing, and $q$ is the order of the moving average part. Once we have the desired forecast $y_t'$, we can undifference the forecast values to obtain the forecast of the original time-series. We denote such a model by *ARIMA(p,d,q)*.

Auto ARIMA models can be implemented using the auto.arima() function in the forecast R package. On the website, ARIMA models are used to predict is used to implement ARIMA time-series prediction on the number of daily new cases. For every given time-series, the script automatically estimates the parameters in the ARIMA fitting model and finds the best ARIMA model to the data based on the AIC score. AIC is an abbreviation of the Akaike Information Criterion, and it estimates the quality of a mode fit l relative to a collection of data models. Specifically, AIC estimates the estimator of out-of-sample prediction error, and so smaller values are desirable. Given a particular statistical data model with $k$ estimated parameters number and $\hat{L}$ the maximum likelihood function value. Then the model's AIC value is given by

$$AIC = 2k - 2\ln(\hat{L}). \qquad (5)$$

Thus, for ARIMA models, AIC can be computed as follows,

$$AIC = -2\log L + 2(p + q + k). \qquad (6)$$

For every given time-series, auto.arima() chooses the parameters which give the smallest AIC and forecasts the values for the next n days.[9] For the forecast displayed on the website, we use n = 5 days. As a part of the output from the auto.arima() function, the 95% intervals are taken to plot the transparent orange ribbon around the mean prediction shown in Figures 9 and 10.

## 5- Feature 3: Common Symptoms

### *5-1- Introduction*

This subsection discusses common symptoms experienced by COVID-19 patients and the information can be obtained from the website. The section Common Symptoms contains an interactive visual summary of the most common symptoms associated with the disease (Figure 11). The information discussed in this section is derived from medical records of infected patients around the world made publicly available by the Open COVID-19 Data Working Group in their nCoV2019 data repository. Due to data quality issues and the uncertain nature of the disease, it is difficult to estimate the true prevalence of the symptoms among infected patients. Hence, their prevalence measure is standardized to a 0-to-10 scale and is represented by the horizontal axis of the plot.

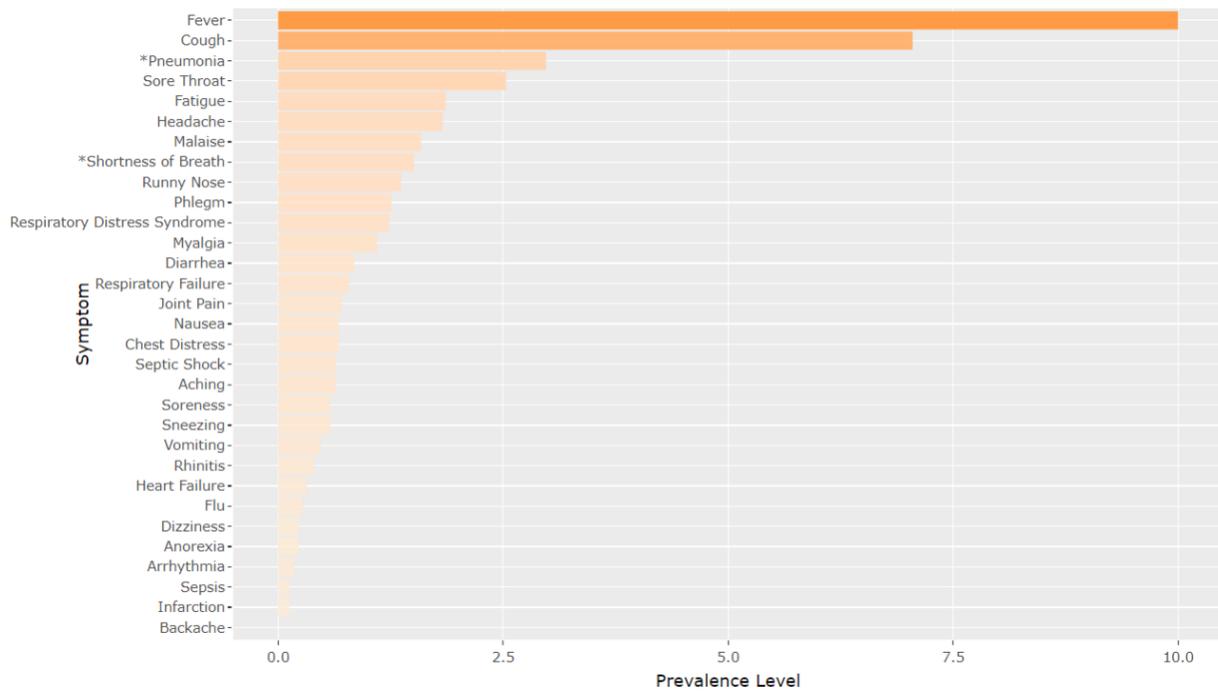

Figure 11. Common Symptoms of COVID-19.

## *5-2- N-Gram Tokenization*

Since the symptom variable from the patient-level data contains descriptive sentences of a patient's symptoms (e.g. "Moderate fever 38.5ºC, cough, strong headache"), we have to apply natural language processing techniques such as n-gram tokenization to transform and preprocess the data. The goal is to convert the descriptive sentences into a set of binary indicator variables, as shown by the simple example in Figure 12.

| Patient | Symptom |
|---|---|
| 1 | Moderate fever 38.5oC, cough, strong headache |
| 2 | Fever, pneumonia, fatigue |
| 3 | Fever, headache, fatigue |

| Patient | Fever | Cough | Headache | Pneumoni | Fatigue |
|---|---|---|---|---|---|
| 1 | 1 | 1 | 1 | 0 | 0 |
| 2 | 1 | 0 | 0 | 1 | 1 |
| 3 | 1 | 0 | 1 | 0 | 1 |

Figure 12. Example of Converting Sentences to Binary Indicators.

The process of word tokenization refers to splitting a sample of text into words or phrases. In addition, n-gram tokenization refers to tokenization that splits the text into phrases which contain n words. For example, unigram tokenization turns the sentence "*he has shortness of breath*" into *[he, has, shortness, of, breath]* while trigram tokenization turns the sentence into *[he has shortness, has shortness of, shortness of breath]*.

As an attempt to collect all of the recorded symptoms in the dataset, we apply n-gram tokenization to every descriptive sentence and compute the frequency of each token for n = {1, 2, 3, 4}. As anticipated, we can obtain a list of the most common symptoms from the symptom records by looking through the processed output from n-gram tokenization (Figure 13).

| unigram | bigram | trigram | quadgram |
|---|---|---|---|
| aching/aches | chest distress | acute respiratory distress | acute respiratory distress syndrome |
| anhelation | chest tightness | shortness of breath | |
| anorexia | heart failure | acute respiratory failure | |
| arrhythmia | joint pain | | |
| backache | runny nose | | |
| coriza | septic shock | | |
| cough | sore throat | | |
| ... | | | |

Figure 13. Examples of N-Gram Output.

After obtaining a comprehensive list of symptoms, we create a dictionary of phrases for each symptom and loop through all descriptive sentences to see if they contain any phrase in any dictionary. For example, the dictionary for cough is *[cough, coughing]*, and any sentence that contains *cough* or *coughing* will take the value of 1 for cough's binary indicator variable and 0 if otherwise. By the end of the loop, we would have converted the descriptive sentences into a set of binary indicator variables in the format shown in Figure 12.

*5-3- Min-Max Normalization*

After the application of n-gram tokenization to create all the necessary binary indicator variables, we can then obtain the aggregated count of patients for every symptom by calculating the numerical sum of each binary indicator variable. To better communicate the level of prevalence of each symptom, we apply min-max normalization to the columnar sums to standardize each data point on a scale of 0 to 10. For any symptom's columnar sum, $S_i$, its scaled value is computed as follows,

$$S_{i,scaled} = \frac{S_i - S_{\min}}{S_{\max} - S_{\min}} \cdot 10, \quad (7)$$

where *i* refers to the symptom index *i* on the vertical axis in Figure 11. The scaled value is an abstract representation of the symptom's prevalence relative to other symptoms, and it does not reflect the true prevalence of the symptom among patients who have been infected with COVID-19.

*5-4- Logistic Regression*

A logistic regression model is built to identify risk factors that could potentially increase a patient's likelihood of dying from COVID-19, and more generally for any model with a binary outcome. Once we have formed all the binary indicator variables for symptoms, we use them along with other variables as predictors to build a logistic regression model to predict a patient's binary outcome, such as whether the patient recovered from the disease or died from the disease. We next review the logistic regression model and discuss hypothesis testing of the model's coefficients.

Let y be a binary output variable, taking on values 0 or 1, where 1 denotes the patient's death and 0 otherwise. Given x is the vector the covariates of $E(y|x)$, by

$$P(y=1 \mid x, \beta) = \frac{e^{\beta^T x}}{1 + e^{-\beta^T x}} \text{ and } P(y=0 \mid x, \beta) = \frac{1}{1 + e^{\beta^T x}}. \tag{8}$$

We invert the transformation and obtain the logit function,

$$g(x \mid \beta) = \log\left(\frac{P(y=1 \mid x, \beta)}{1 - P(y=1 \mid x, \beta)}\right) = \beta^T x. \tag{9}$$

If we apply a logistic regression model to the dataset with n observations, $D = \{(x^1, y^1), ..., (x^n, y^n)\}$, the condition likelihood of a single data observation is

$$P(y^i \mid x^i, \beta) = P(y^i = 1 \mid x^i, \beta)^{y^i} P(y^i = 0 \mid x^i, \beta)^{1-y^i},$$

$$\text{where } P(y^i = 1 \mid x^i, \beta) = \frac{e^{\beta^T x^i}}{1 + e^{-\beta^T x^i}} \text{ and } P(y^i = 0 \mid x^i, \beta) = \frac{1}{1 + e^{\beta^T x^i}}. \tag{10}$$

This gives the total log-likelihood

$$l(\beta \mid X, Y) = \sum_{i=1}^{n} y^i \log(P(y^i = 1 \mid x^i, \beta)) + (1 - y^i) \log(P(y^i = 0 \mid x^i, \beta)).$$

To find the maximum likelihood estimators of $\beta$, we differentiate the above expression with respect to each $\beta$ components and set them equal to 0 to find the solutions:

$$\frac{\partial l(\beta \mid X, Y)}{\partial \beta_k} = \sum_{i=1}^{n} y^i \frac{1}{P(y^i = 1 \mid x^i, \beta)} \frac{\partial P(y^i = 1 \mid x^i, \beta)}{\partial \beta_k} - (1 - y^i) \frac{1}{P(y^i = 0 \mid x^i, \beta)} \frac{\partial P(y^i = 0 \mid x^i, \beta)}{\partial \beta_k}$$

$$\frac{\partial l(\beta \mid X, Y)}{\partial \beta_k} = \sum_{i=1}^{n} x_i^k (y^i - P(y^i = 1 \mid x^i, \beta)) \tag{11}$$

Since there is no closed-form solution to these equations, they must be solved iteratively using a numerical method, such as the Newton-Raphson method. Assuming that we have successfully estimated all the coefficients, $\hat{\beta}$, using a numerical method, we then conduct hypothesis testing to evaluate if the predictors have statistically significant associations with the outcome variable.[12]

Using large sample theory, we apply the Wald test to test any selected coefficient in the model. If the $j^{th}$ coefficient is of interest, null and alternative hypotheses are

$$H_0: \beta_j = 0$$
$$H_1: \beta_j \neq 0.$$

To calculate the test statistics associated with the coefficient, we compute z as follows,

$$z = \frac{\hat{\beta}_j - \beta_{j0}}{SE(\hat{\beta})} = \frac{\hat{\beta}}{SE(\hat{\beta})}, \tag{12}$$

which has a standard normal distribution under the null hypothesis. Here $\hat{\beta}$ is the estimated coefficient and its standard deviation, $SE(\hat{\beta})$, is calculated by taking the inverse of the estimated information matrix. Extension to the case when we want to test multiple coefficients is available but not discussed here.

We now use a multiple logistic model to fit the COVID-19 patient dataset and include symptoms and demographic variables as predictors, we have the following logit function,

$$\log it(\pi_i) = \beta_0 + \beta_1 age + \beta_2 sex.female + \beta_3 chronic.disease.1 + \beta_4 respiratory.distres.syndrome.1 +$$
$$\beta_5 respiratory.failure.1 + \beta_6 chest.distress.1 + \beta_7 shortness.of.breath.1 + \beta_8 heart.failure.1 +$$
$$\beta_9 runny.nose.1 + \beta_{10} spetic.shock.1 + \beta_{11} sore.throat.1 + \beta_{12} anorexia.1 + \beta_{13} arrhythmia.1 +$$
$$\beta_{14} cough.1 + \beta_{15} diarrhea.1 + \beta_{16} dizziness.1 + \beta_{17} fatigue.1 + \beta_{18} fever.1 + \beta_{19} headache.1 +$$
$$\beta_{20} \inf arction.1 + \beta_{21} malaise.1 + \beta_{22} mya \lg ial.1 + \beta_{23} phlegm.1 + \beta_{24} pneumonia.1 +$$
$$\beta_{25} sepsis.1 + \beta_{26} soreness.1$$

(13)

where π is the probability that a patient will die from COVID-19 or not. The above symptoms are taken from Figure 11 and the indicator variable takes on the value 1 if the patient has that particular symptom and 0 otherwise. After the model is fitted, we need to assess the quality of the model fit. We use cross-validation to evaluate the model's ability to predict future or out-of-sample responses using various goodness of fit measures.[13] We also apply model diagnostic tools in order to flag potential problems such as overfitting or selection bias. These additional analyses provide insights on the model's level of robustness and generalization of the new data that is not a part of its training data. One round of cross-validation partitions a sample of data into complementary subsets, performs the analysis on one subset (i.e. the training set), and validates the analysis on the other subset (i.e. the validation set or testing set). To reduce variability, we perform this procedure k times by initially partitioning the data into *k* subsets. Figure 14 demonstrates a visual summary of the process when *k* = 5.

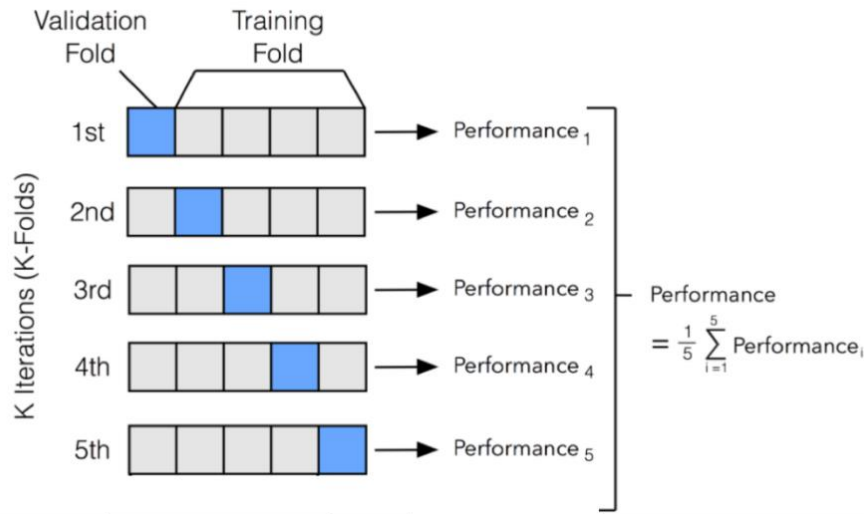

Figure 14. 5-Fold Cross Validation Visualization.

Using the *caret* and *glmnet* packages in R, we use *glmnet* to create our logistic regression model and feed it into the cross-validation function from caret. We then perform 5-fold cross-validation to compute the overall accuracy and the ROC curve of the logistic regression model. Our results show that the model has an overall accuracy of 0.900 with a standard deviation of 0.03, and the ROC curve is shown in Figure 16. We recall ROC is short for receiver operating characteristic curve, and it is a graphical plot that illustrates the diagnostic ability of a binary classifier system as its discrimination threshold is varied. AUROC is short for the area under the ROC curve and a random classifier has a baseline AUROC of 0.5. In general, a binary classifier is more desirable if it has a larger AUROC value.

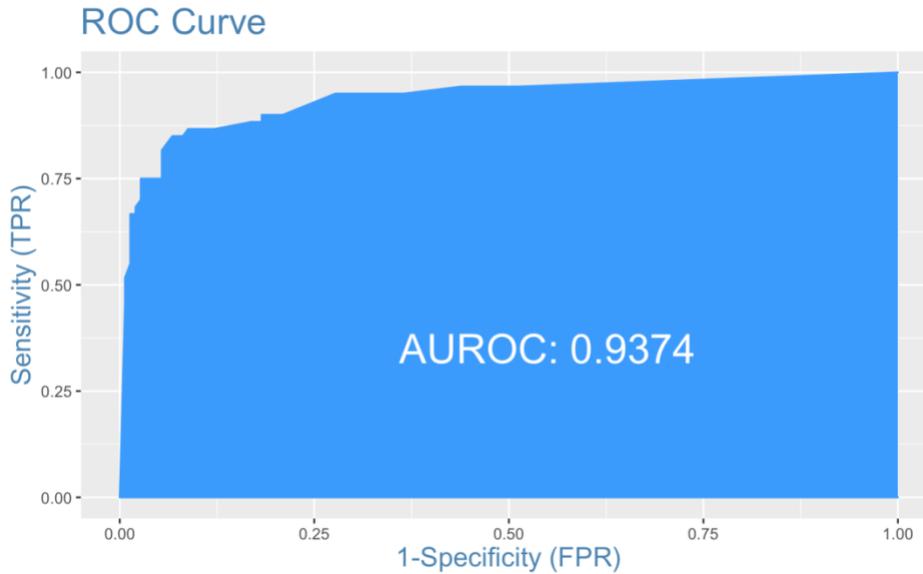

Figure 15. ROC Curve.

After confirming the quality of the model, we apply the same model to the whole dataset, and the table below displays our results for the fitted model.

| Variable | Estimate | Std. Error | z value | Pr(>|z|) |
|---|---|---|---|---|
| (Intercept) | -8.10 | 0.73 | -11.11 | **<0.001** |
| age | 0.11 | 0.01 | 10.09 | **<0.001** |
| sexFemale | -0.37 | 0.30 | -1.23 | 0.22 |
| chronic_disease_binary1 | 0.51 | 0.42 | 1.23 | 0.22 |
| respiratory_distress_syndrome1 | 19.53 | 1402.78 | 0.01 | 0.99 |
| respiratory_failure1 | 19.46 | 1832.13 | 0.01 | 0.99 |
| chest_distress1 | 18.38 | 4619.90 | 0.00 | 1.00 |
| shortness_of_breath1 | 2.56 | 1.11 | 2.30 | **0.02** |
| heart_failure1 | 17.76 | 3812.91 | 0.00 | 1.00 |
| runny_nose1 | -17.40 | 3196.51 | -0.01 | 1.00 |
| septic_shock1 | 16.09 | 2113.25 | 0.01 | 0.99 |
| sore_throat1 | -0.04 | 1.15 | -0.04 | 0.97 |
| anorexia1 | 17.33 | 7604.24 | 0.00 | 1.00 |
| arrhythmia1 | 12.15 | 2892.88 | 0.00 | 1.00 |
| cough1 | 0.69 | 0.60 | 1.15 | 0.25 |
| diarrhea1 | 2.42 | 9.78 | 0.25 | 0.80 |
| dizziness1 | 18.96 | 10754.01 | 0.00 | 1.00 |
| fatigue1 | 1.43 | 1.06 | 1.35 | 0.18 |
| fever1 | 0.79 | 0.50 | 1.59 | 0.11 |
| headache1 | 0.94 | 4.41 | 0.21 | 0.83 |
| infarction1 | 19.96 | 3750.52 | 0.01 | 1.00 |
| malaise1 | -18.27 | 5054.53 | 0.00 | 1.00 |
| myalgia1 | -17.88 | 4579.83 | 0.00 | 1.00 |
| phlegm1 | -15.06 | 4911.26 | 0.00 | 1.00 |
| pneumonia1 | 2.87 | 1.07 | 2.70 | **0.01** |
| sepsis1 | 13.85 | 4111.91 | 0.00 | 1.00 |

Table 1. Logistic Regression Analysis Results.

# 6- Feature 4: Patient Demographics

## *6-1- Introduction*

      This section of the website shows a summary visualization of the distributions of demographic characteristics of patients with recorded demographic information (Figure 16). The demographic information available from the data source includes patient age and gender.

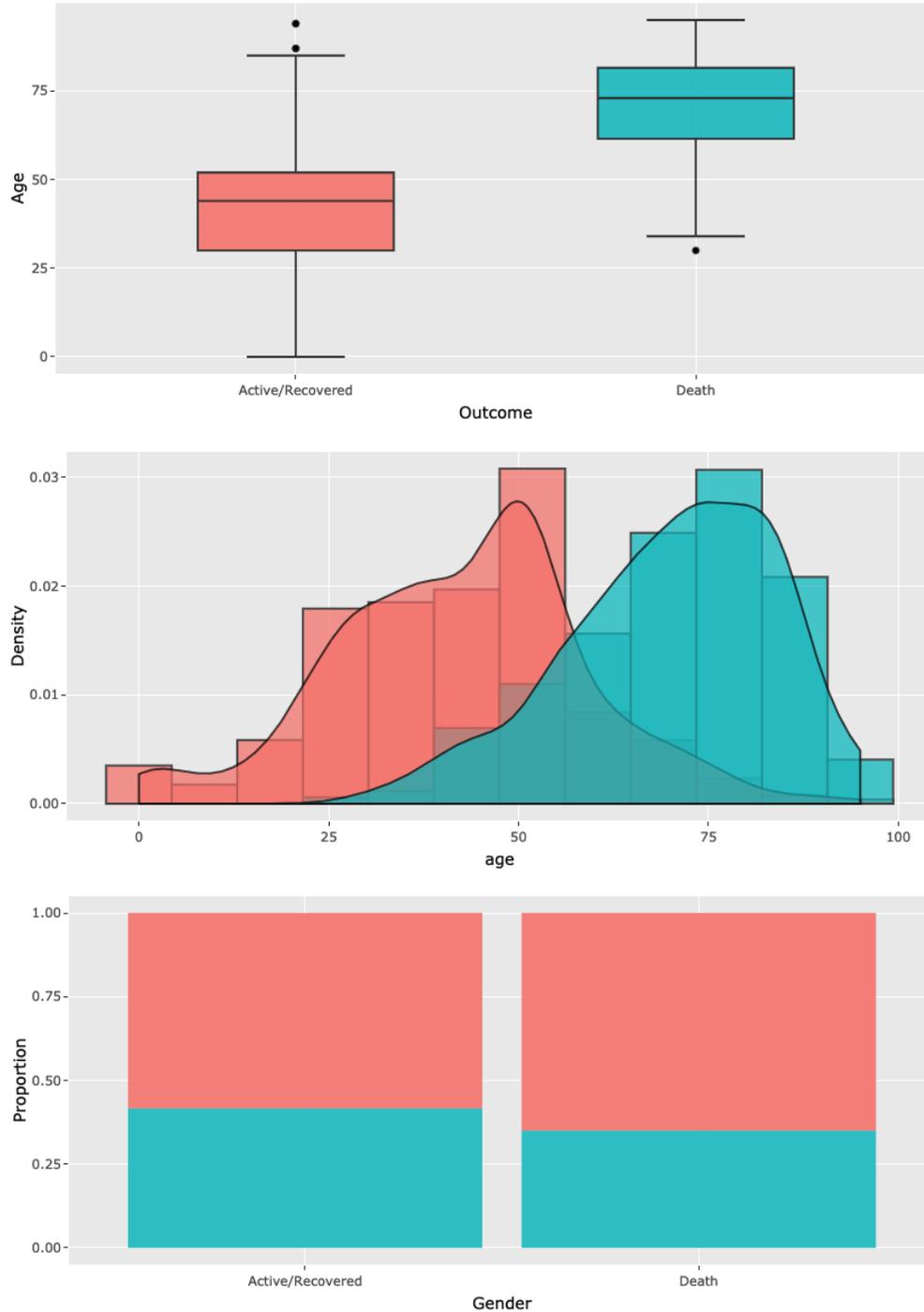

Figure 16. Summary Visualizations of Demographics Characteristics of COVID-19 Patients.

## 6-2- Two Sample T-Test

To determine if there is a statistically significant difference in the age of two patient groups, active/recovered or death, we conduct a two-sample t-test. Two-sample t-test is a hypothesis testing method to compare two continuous-data distributions. More precisely, it tests to determine if the means of two continuous distributions are equal. The assumptions of the two-sample t-test properly are

i. The data are continuous (not discrete),
ii. The data follow the normal probability distribution,
iii. The variances of the two populations are equal,
iv. The two samples are independent. There is no relationship between the individuals in one sample as compared to the other,
v. Both samples are simple random samples from their respective populations and each individual in the population has an equal probability of being selected in the sample.[14]

Assumption (i) is satisfied as the value of age is continuous. However, assumptions (iv) and (v) may not be valid due to potential data quality issues such as missing data. We presume they are satisfied and proceed with cautions. For assumptions (ii), we validate the data's normality using QQ plots as shown in Figure 17.

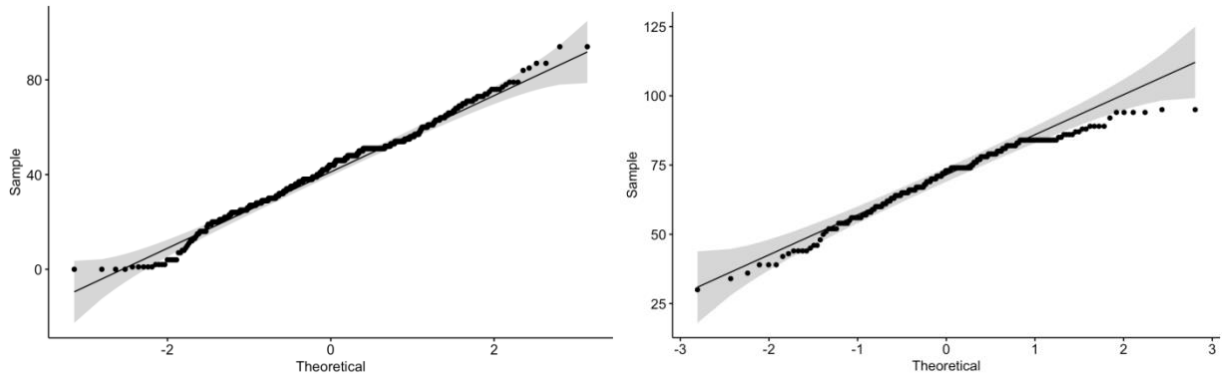

Figure 17. QQ Plots of Ages of Active/Recovered Patients (Left) and Dead Patients (Right).

The data points appear to be decently consistent with the quantiles of a normal distribution. For assumption (iii), we apply the F-test to test the equality of variances for the two groups and the hypotheses are

$$H_0 : \sigma_X = \sigma_Y$$
$$H_1 : \sigma_X \neq \sigma_Y.$$

The test statistics is

$$F = \frac{S_X^2}{S_Y^2} = 1.4895, \tag{14}$$

where $S_X$ denotes the sample standard deviation of the age of active/recovered patients, $S_Y$ denotes the sample standard deviation of the age of dead patients. Under the null hypothesis, F has an F-distribution with n and m degrees of freedom and for our data, the corresponding p-value is 0.00097. Thus, we have sufficient evidence to reject the null hypothesis and conclude that the variances of the two groups are unequal at the alpha level of 0.05.

Since the sample variances have been shown to be unequal, we use a two sample t-test with un-pooled variances to test whether the means from the two groups are equal, as follows,

$$H_0 : \mu_X = \mu_Y$$
$$H_1 : \mu_X \neq \mu_Y$$
$$t = \frac{\overline{X} - \overline{Y}}{\sqrt{\frac{S_X^2}{n} + \frac{S_Y^2}{m}}}, \quad (15)$$

$$df = v = \frac{(\frac{S_X^2}{n} + \frac{S_Y^2}{m})^2}{\frac{(\frac{S_X^2}{n})^2}{n-1} + \frac{(\frac{S_Y^2}{m})^2}{m-1}}. \quad (16)$$

Under the null hypothesis, t has a student-t distribution with the degree of freedom of v. With a test statistic of -23.785 and a p-value that is approximately 0, we reject the null hypothesis at the alpha level of 0.05. Hence, we conclude that the average age of patients who are active or recovered is different from the average of patients who have died from COVID-19.

*6-3- Chi-Square Test*

To determine if there is a statistically significant association between a patient's gender and a patient's outcome, we conduct a Chi-Square test for a test of association of a 2x2 contingency table. The assumptions of the Chi-Square Test are

i. The data are continuous (not discrete),
ii. The data follow the normal probability distribution,
iii. The variances of the two populations are equal,
iv. The two samples are independent. There is no relationship between the individuals in one sample as compared to the other,
v. Both samples are simple random samples from their respective populations and each individual in the population has an equal probability of being selected in the sample.[15]

Assumptions (i) and (ii) are met since we are observing counts of patients who are either male or female, and either active/recovered or deceased. In addition, assumption (iv) is satisfied as shown by the 2x2 contingency table below. We presume assumption (iii) to hold and proceed.

After filtering out missing data to create a subset of patient data with recorded genders and outcomes, we obtained the following 2x2 contingency table,

|        | Active/Recovered | Death |
|--------|------------------|-------|
| Male   | 299              | 132   |
| Female | 213              | 71    |

Table 2. The 2x2 Contingency Table

More generally, in a *rxc* contingency table, let $R_i$ and $C_j$ be the row sum of row *i* and the column sum of column *j* respectively, and let *n* be the total in the sample. We calculate the Chi-square test statistic as follows,

$$X^2 = \sum_{i=1}^{r}\sum_{j=1}^{c} \frac{(O_{i,j} - E_{i,j})^2}{E_{i,j}},$$
where $E_{i,j} = (R_i \cdot C_j)/n.$ (17)

Under the null hypothesis, gender has no association with whether the COVID-19 patients died from the disease or not, $X^2$ has a Chi-square distribution with the degree of freedom of $(r-1)(c-1)$, where $r$ and $c$ are the number of rows and columns in the contingency table, respectively. For our data in Table 2, we obtained a test statistic of 2.6657 and a p-value of 0.1025. The result does not provide sufficient evidence for rejecting the null hypothesis at the alpha level of 0.05 and we conclude that a patient's gender has no statistically significant association with whether the COVID-19 patients died from the disease or not.

## 7- Conclusion

This research describes a web-based application for assessing real-time data for analyzing the latest trends of COVID-19 across different regions, COVID-19's symptoms and patient demographics. The research also highlights details of the methodologies behind the real-time COVID-19 tracker website, which include automatic data ingestion pipeline, data transformation and normalization, time-series forecast with ARIMA model forecast, text mining techniques, and logistic regression model. The literature also explains how these methodologies are combined to produce predictions and insights. The unique and innovative components of the analytical approach in this paper include the automatic data ingestion and processing associated with the website, as well as the NLP-oriented approach to discover the common symptoms. However, we need to be cautious about accepting the conclusions as there are potential data quality issues such as when the patient-level data has a substantial amount of missing data and erroneous entries. To verify the findings in this research, we should reproduce our findings in the research when we have access to an updated dataset repeatedly and towards the end of the pandemic.

During a global-level pandemic such as COVID-19, it is paramount for the public to have access to the latest status of the outbreak and be well-informed of relevant insights into the disease. A platform such as a real-time COVID-19 tracker website will assist the public community to disseminate accurate and reliable insights into the spread of COVID-19. The research and effort behind this project are motivated by the social responsibility to spread awareness to the common public by providing scientific-based data analysis, prediction, and relevant findings. This paper and research project are still ongoing research as many more investigations regarding COVID-19 can be carried out. It will serve as an initial step to unravel the many uncertainties that revolve around this global pandemic.

## 8- Declarations

### 8-1- Data Availability Statement

The data presented in this study are available on request from the corresponding author.

### 8-2- Funding

The author(s) received no financial support for the research, authorship, and/or publication of this article.

### 8-3- Acknowledgements

The data used in this paper are publicly available on the CSSE data repository by Johns Hopkins University (https://github.com/CSSEGISandData/COVID-19) and the nCoV2019 data repository by the Open COVID-19 Data Working Group (https://github.com/beoutbreakprepared/nCoV2019/tree/master/latest_data).

*8-4- Conflicts of Interest*

The author declares that there is no conflict of interests regarding the publication of this manuscript. In addition, the ethical issues, including plagiarism, informed consent, misconduct, data fabrication and/or falsification, double publication and/or submission, and redundancies have been completely observed by the authors.

# 10- References